\documentclass[]{spie}  

\usepackage{amsmath,amsfonts,amssymb}
\usepackage{graphicx}
\usepackage[colorlinks=true, allcolors=blue]{hyperref}

\usepackage{latexsym}

\usepackage{color, soul}
\usepackage{xargs}
\usepackage{tikz}
\usepackage{todonotes}
\usepackage{multirow} 
\usepackage{verbatim}
\usepackage{nameref,hyperref}

\usepackage{multicol}
\usepackage{multirow}
\usepackage{hyperref}
\usepackage{subcaption}
\usepackage{bm}
\usepackage{booktabs}
\usepackage{enumitem}

\newlength\savewidth\newcommand\shline{\noalign{\global\savewidth\arrayrulewidth
  \global\arrayrulewidth 1.5pt}\hline\noalign{\global\arrayrulewidth\savewidth}}
  
\title{Random forest-based out-of-distribution detection for robust lung cancer segmentation}

\author[]{Aneesh Rangnekar}
\author[]{Harini Veeraraghavan}
\affil[]{Department of Medical Physics, Memorial Sloan Kettering Cancer Center, New York, USA}

\authorinfo{Send correspondence to Aneesh Rangnekar (Email: \href{mailto:rangnea@mskcc.org}{rangnea@mskcc.org})}

\begin{document} 
\maketitle

\begin{abstract}
Accurate detection and segmentation of cancerous lesions from computed tomography (CT) scans is essential for automated treatment planning and cancer treatment response assessment. Transformer-based models with self-supervised pretraining can produce reliably accurate segmentation from in-distribution (ID) data but degrade when applied to out-of-distribution (OOD) datasets. We address this challenge with RF-Deep, a random forest classifier that utilizes deep features from a pretrained transformer encoder of the segmentation model to detect OOD scans and enhance segmentation reliability. The segmentation model comprises a Swin Transformer encoder, pretrained with masked image modeling (SimMIM) on 10,432 unlabeled 3D CT scans covering cancerous and non-cancerous conditions, with a convolution decoder, trained to segment lung cancers in 317 3D scans. Independent testing was performed on 603 3D CT public datasets that included one ID dataset and four OOD datasets comprising chest CTs with pulmonary embolism (PE) and COVID-19, and abdominal CTs with kidney cancers and healthy volunteers. RF-Deep detected OOD cases with a FPR95 of 18.26\%, 27.66\%, and $<$ 1.0\% on PE, COVID-19, and abdominal CTs, consistently outperforming established OOD approaches. The RF-Deep classifier provides a simple and effective approach to enhance reliability of cancer segmentation in ID and OOD scenarios.
\end{abstract}


\section{Purpose}
Pretrained transformer-based deep learning (DL) methods combined with convolutional decoders have demonstrated capability to segment organs and tumors from radiological images~\cite{Willemink2022,jiang2022self,NguyenAAAI2023,YanWACV2023,Qayyum2023BHI,gu2024build}. A major challenge in deploying DL segmentation models at scale in research and clinical settings is the potential accuracy degradation when the same models are applied to real-world scenarios that differ from those seen during training. For example, models trained to segment malignant lung cancers from chest CTs for treatment planning may also be applied to scans with benign nodules from lung cancer screening or scans containing unrelated diseases such as pulmonary embolisms. Although all of these involve chest CTs, models trained for cancer segmentation can produce unanticipated and incorrect results when applied outside their intended scope. 

Traditional segmentation evaluation metrics such as the Dice similarity coefficient (DSC) and Hausdorff distance at 95th percentile (HD95) are designed to assess model performance on in-distribution (ID) datasets, but are insufficient indicators of the model's robustness to out-of-distribution (OOD) data. OOD detection approaches relying on model confidence scores, such as MaxSoftmax~\cite{hendrycks17baseline}, often fail in scenarios where models produce confidently incorrect segmentations, occurring in both medical~\cite{yeung2023calibrating} and natural image analyses~\cite{hendrycks2019anomalyseg}. These methods have been applied to distinctly different disease sites such as the lung and abdomen~\cite{mehrtash2020confidence,zimmerer2022mood}, a relatively easier task compared to detecting OOD cases within the same site. Alternative approaches using secondary models, such as VQ-GANs~\cite{graham2022transformer,pinaya2022unsupervised}, require substantially large secondary training data, are computationally intensive, and lack interpretability, limiting their practical utility in high-throughput clinical workflows. Finally, radiomics feature-based methods that leverage standardized  features have been shown to be effective at distinguishing scans of organs but have not generalized to tumors~\cite{vasiliuk2023limitations,konz2024radmetricmedicalimage}. To address these limitations, we developed a random forest classifier combining deep features (RF-Deep) of a model trained to segment lung cancers to detect OOD scans. 

We evaluated our OOD detection approach on both, far-OOD scans consisting of diseases occurring in different anatomic sites, and the more challenging near-OOD scans where different diseases occur in the same anatomic site with similar appearance statistics as the training datasets. In summary, our contributions are: \textbf{(a)} a lightweight, random forest classifier based OOD detection framework utilizing deep features trained to extract useful feature representations for ID cases. The deep features use the encoder features in order to leverage the robustness of the pretrained encoder to imaging distribution variations inherent even in ID datasets.  The classifier is trained on a small set of ID and representative OOD examples, following the outlier exposure paradigm~\cite{hendrycks2018deep}, \textbf{(b)} a demonstration of the effectiveness of our approach on lung cancer segmentation from 3D CT scans, including both near-OOD scenarios in chest CTs (pulmonary embolism and COVID-19) and far-OOD scenarios in abdominal CTs (kidney cancers and non-cancerous pancreas), and \textbf{(c)} a systematic comparison of our RF-Deep approach against common OOD detection methods (like MaxSoftmax) and a radiomics feature–based classifier (RF-Radiomics), using five public datasets totaling 603 patient scans.

\begin{figure}[t]
    \centering
    \includegraphics[width=0.9\linewidth]{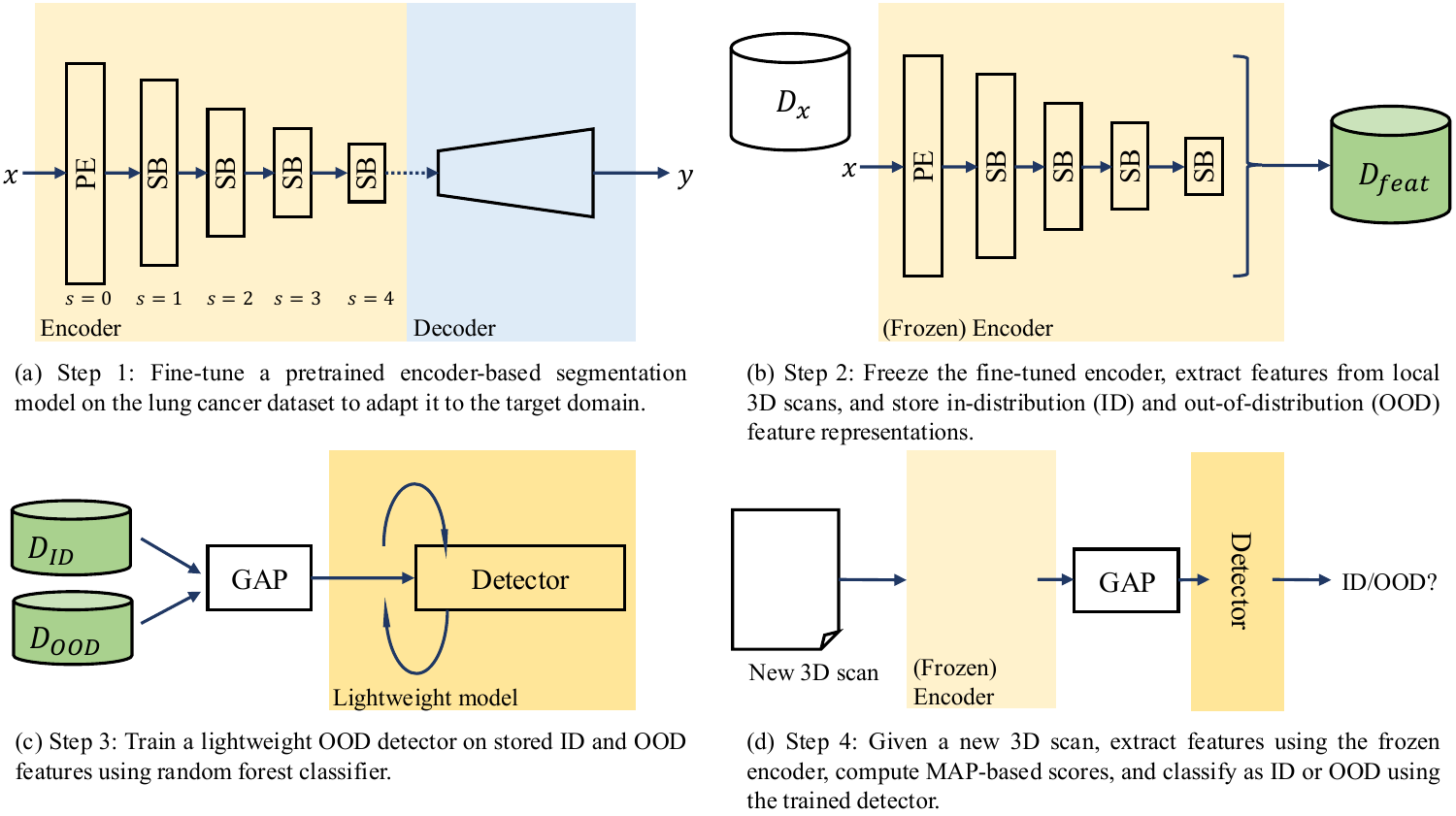}
    \caption{High-level pipeline of the proposed out-of-distribution (OOD) detection framework, focusing on segmentation for lung cancer CT scans. After fine-tuning, the encoder is frozen, and features are extracted from the patch embedding (PE) layer and Swin blocks (SB) across all four stages to be used with a random forest classifier.}
    \label{fig:summary}
\end{figure}

\section{Method}

\textbf{OOD detection task definition.} Let $\mathcal{D}_\text{in}$ denote the distribution of lung cancer CT scans, closely matching the training dataset used to create the lung tumor segmentation model, and $\mathcal{D}_\text{out}$ represents the scans that differ in pathology and anatomic site. The objective is to determine whether a new scan $x$ belongs to $\mathcal{D}_\text{in}$ or $\mathcal{D}_\text{out}$, at the scan level, via a scoring mechanism $S(x)$. Since tumor segmentation is binary, aggregation is restricted to the predicted tumor regions, leveraging model-relevant spatial context while excluding irrelevant anatomy.

\noindent \textbf{Segmentation model architecture.} A hybrid transformer encoder-convolutional decoder architecture was employed. The encoder extracts global and spatially local contexts through a hierarchical Swin Transformer~\cite{liu2021swin} encoder. The decoder leverages local spatial precision of a U-Net–based~\cite{ronneberger2015u} convolutional network to accurately delineate anatomical boundaries. The encoder uses a depth configuration of $2-2-12-2$ across four stages with $4 \times 4 \times 4$ patch size, enabling multi-scale feature aggregation, while windowed self-attention reduces computational complexity yet preserves global context for large volumetric inputs (128 $\times$ 128 $\times$ 128 voxels). Encoder was initialized with pretrained weights with self-supervised learning performed using SimMIM~\cite{xie2022simmim} method that uses masked image modeling (MIM) approach to predict and reconstruct masked image patches. Pretraining used 10,432 unlabeled 3D CT scans sourced from public and institutional datasets~\cite{jiang2022self,jiang2024self} with MIM task performed by randomly masking 75\% of 3D patches in the image. Note that SSL pretraining is an unsupervised pretraining approach that does not require labeled image datasets for pretraining. Pretrained encoder was used to extract features that are robust to CT image acquisition features to accurately identify ID cases despite imaging differences and detect OOD cases.

\noindent\textbf{OOD Detection with RF-Deep classifier.} Unlike object-centric photographic images, medical images may contain varying anatomic extent, making a global image-based OOD assessment inefficient and potentially less accurate. Hence, we leveraged the target task-relevant regions or tumors as extracted by the segmentation model to focus OOD detection on relevant image regions. Our approach (Fig.~\ref{fig:summary}) consists of four steps: \textbf{(i)} fine-tune a segmentation model for lung tumor segmentation, \textbf{(ii)} obtain tumor-centered 3D image regions, \textbf{(iii)} extract an aggregate feature representation within the generated tumor regions from the multi-scale transformer encoder layers and train an RF-Deep classifier using ID and OOD examples, and \textbf{(iv)} the extracted RF-Deep classifier can then be used to distinguish ID from OOD samples. 

\noindent\textbf{Implementation details.} The segmentation model was trained from a public dataset containing non-small cell lung cancers~\cite{aerts2015data} (N=317). It was implemented using PyTorch~\cite{paszke2019pytorch} and MONAI~\cite{cardoso2022monai}, fine-tuned using cross-entropy and Dice loss with a batch size of 16 across 4 NVIDIA GPUs. A learning rate of $2 \times 10^{-4}$ was used, with linear warm-up and cosine annealing over 1000 epochs. Augmentations included flips, rotations, affine, and intensity shifts. Inputs were normalized (HU $[-400,400]$), resampled (1mm$^3$), and cropped to $128^3$. Sliding window with 50\% overlap was used for inference. RF-Deep classifier consisted of a random forest (1000 trees, max depth 20, balanced weights) trained via the scikit-learn library. Features were extracted from 8 tumor-centered crops per scan, and predictions were averaged at inference. Experiments to detect OOD utilized a held-out ID dataset containing 140 lung cancers was used~\cite{bakr2017data} and four datasets consisting of pulmonary embolism (PE)~\cite{colak2021rsna} (N=120), COVID-19~\cite{ricord2021covid} (N=120), kidney cancers~\cite{heller2023kits21} (N=120), and healthy abdominal CT scans~\cite{roth2015deeporgan} (N=82).

\noindent\textbf{OOD comparison methods.} The RF-Deep classifier was compared against standard OOD methods including MaxSoftmax~\cite{hendrycks17baseline}, MaxLogits~\cite{hendrycks2022logits}, energy~\cite{liu2020energy}, and entropy measures. In addition, a RF-radiomics classifier was created using 293 IBSI-compliant radiomic features~\cite{zwanenburg2020image} extracted from detected tumor regions using PyCERR radiomics library~\cite{deasy2003cerr,apte2018extension}. In order to prevent accuracy degradation from correlated features, recursive feature elimination was employed. RF used 1000 trees, max depth of 20, and balanced weights.  

\noindent\textbf{OOD experiment protocol.} RF-based approaches used a fixed 40/60 patient-level train–test split, repeated over 100 seeds. Other baselines used the full cohorts as they require no auxiliary data.

\noindent\textbf{Evaluation metrics.} We report AUROC to measure ID–OOD separability~\cite{davis2006relationship} and FPR95 as a threshold-based metric~\cite{liang2017enhancing}.

\section{Results}

\begin{table}[]
\centering
\def\arraystretch{1.25}
\caption{Out-of-distribution (OOD) detection performance comparing RF-Deep and other methods. Results are reported as AUROC ($\uparrow$) and FPR95 ($\downarrow$), with best values per dataset highlighted in bold.}
\label{tab:my-table}
\resizebox{0.9\textwidth}{!}{%
\begin{tabular}{lllllllll}
\multirow{2}{*}{Method} & \multicolumn{2}{l}{Pulmonary Embolism} & \multicolumn{2}{l}{COVID-19} & \multicolumn{2}{l}{Kidney Cancer} & \multicolumn{2}{l}{Healthy (Pancreas)} \\
 & AUROC ($\uparrow$) & FPR95 ($\downarrow$) & AUROC ($\uparrow$) & FPR95 ($\downarrow$) & AUROC ($\uparrow$) & FPR95 ($\downarrow$) & AUROC ($\uparrow$) & FPR95 ($\downarrow$) \\
 \shline
MaxSoftmax & 88.74 & 37.01 & 89.53 & 53.25 & 96.72 & 12.34 & 97.26 & 14.29 \\
MaxLogits & 90.88 & 34.42 & 91.30 & 37.01 & 97.51 & 7.140 & 98.12 & 6.490 \\
Energy & 90.97 & 35.06 & 91.05 & 37.01 & 97.50 & 7.140 & 98.12 & 6.490 \\
Entropy & 90.59 & 33.81 & 92.47 & 53.24 & 96.97 & 12.23 & 97.55 & 12.95 \\
RF-Radiomics & 88.26 & 40.13 & 90.98 & 36.36 & 95.94 & 19.87 & 95.90 & 15.06 \\
RF-Deep (Ours) & \textbf{95.16} & \textbf{18.26} & \textbf{92.88} & \textbf{27.66} & \textbf{99.81} & \textbf{0.110} & \textbf{99.89} & \textbf{0.720} \\
\bottomrule
\end{tabular}%
}
\end{table}

\begin{figure}[t]
    \centering
    \includegraphics[width=0.9\linewidth]{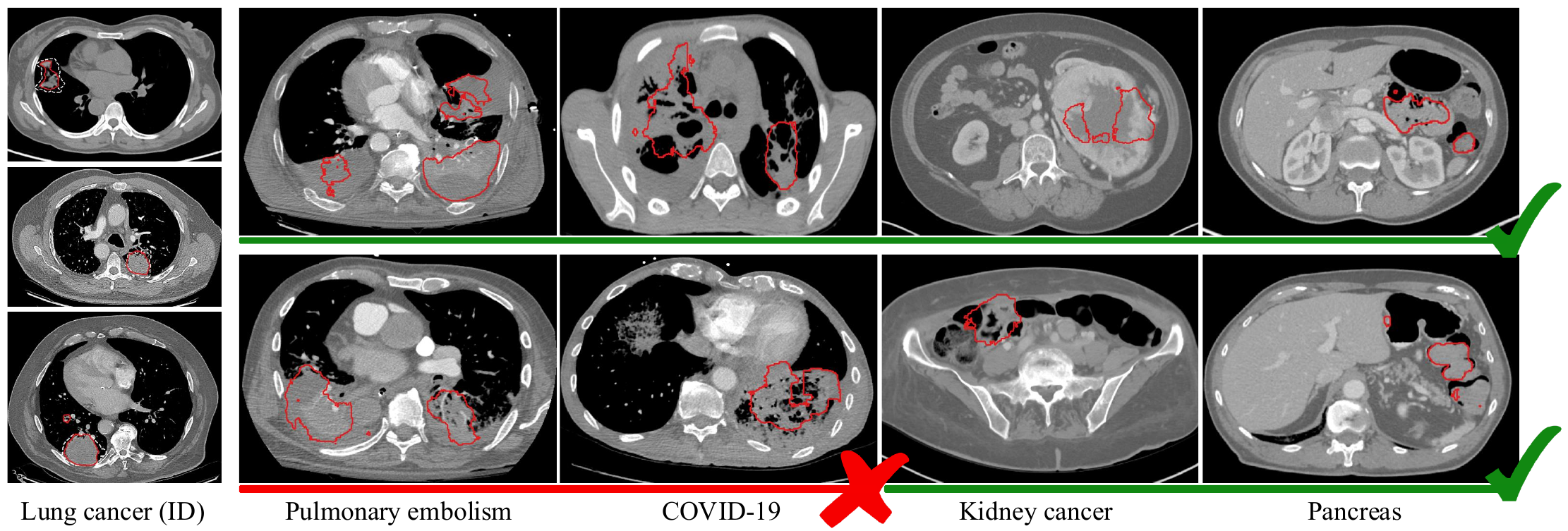}
    \caption{Representative results with green checkmarks indicating correctly detected OOD scans and the red cross highlighting missed OOD detections. The red contours denote segmentations from the model, illustrating that naive confidence measures may fail when tumor-like structures exist in non-cancer cohorts.}
    \label{fig:rf_deep_outputs}
\end{figure}

\begin{figure*}[t]
    \centering
    \begin{subfigure}[t]{0.36\textwidth}
        \centering
        \includegraphics[width=\linewidth]{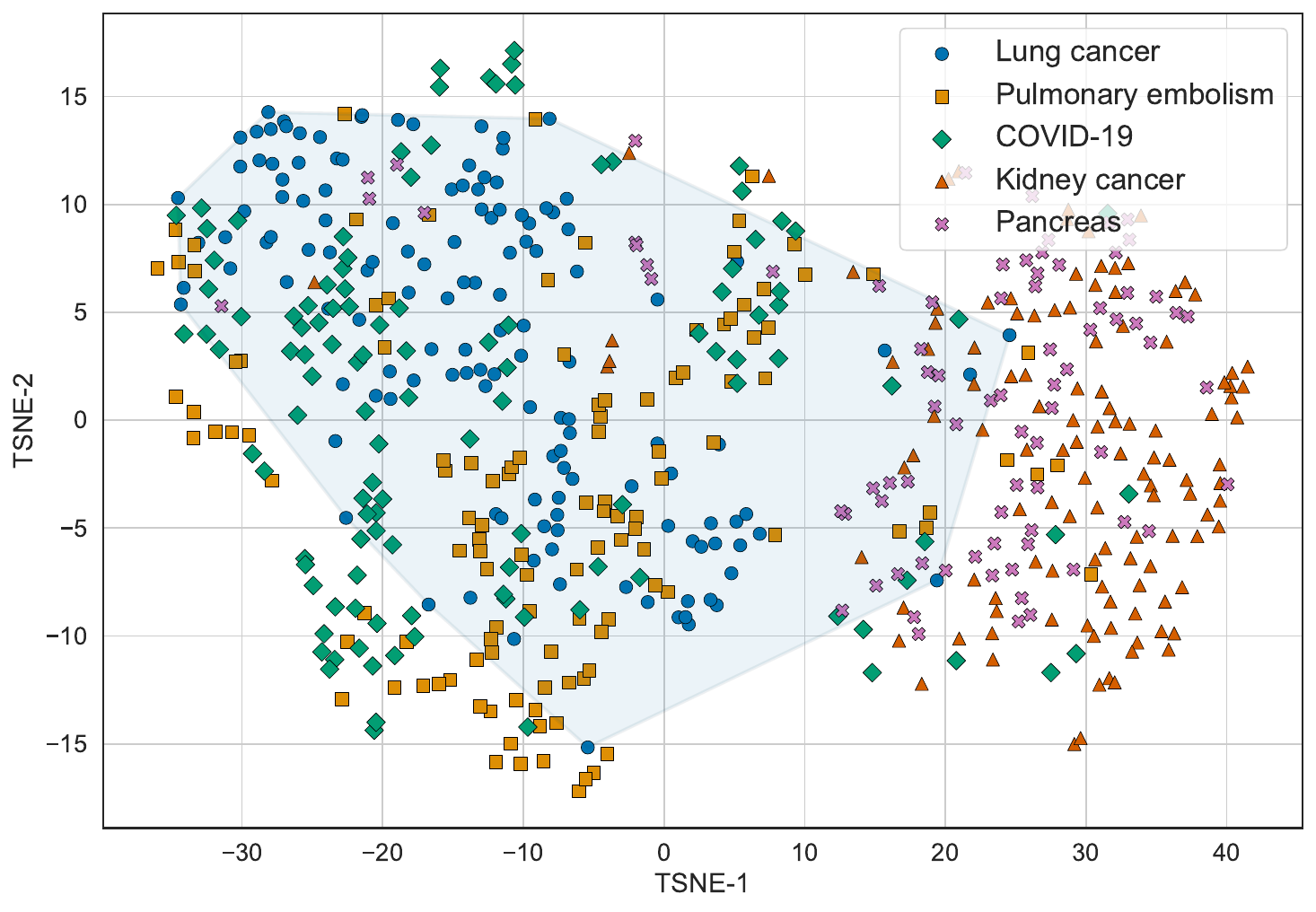}
        \caption{}
        \label{fig:tsne}
    \end{subfigure}
    \hfill
    \begin{subfigure}[t]{0.19\textwidth}
        \centering
        \includegraphics[width=\linewidth]{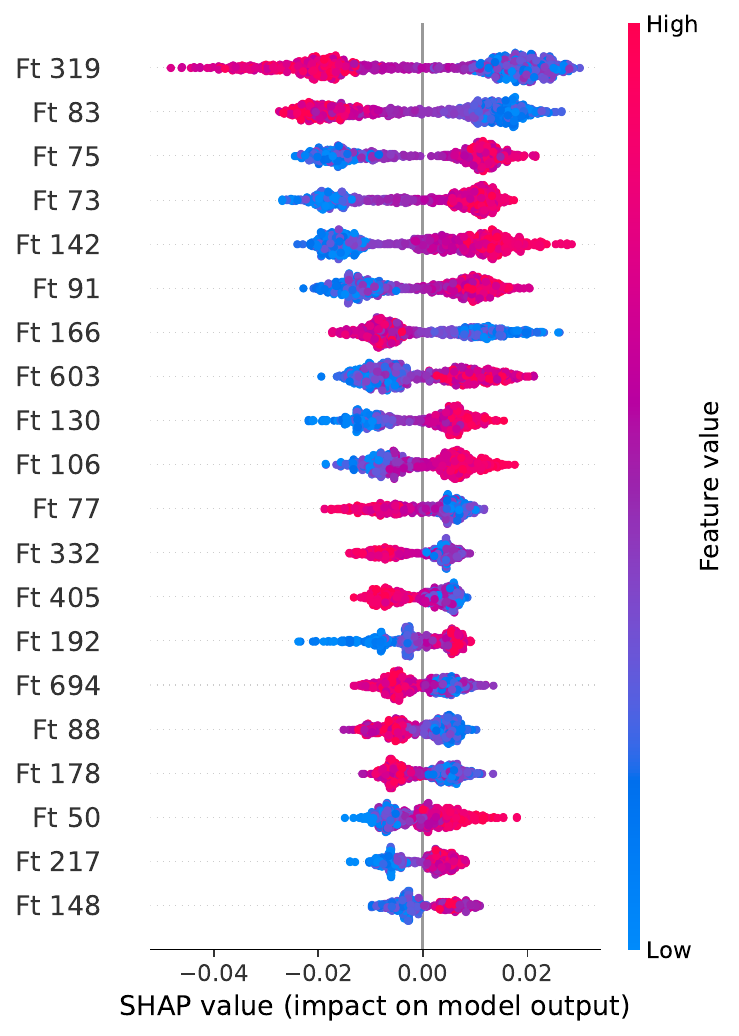}
        \caption{}
        \label{fig:shap_rsna}
    \end{subfigure}
    \hfill
    \begin{subfigure}[t]{0.19\textwidth}
        \centering
        \includegraphics[width=\linewidth]{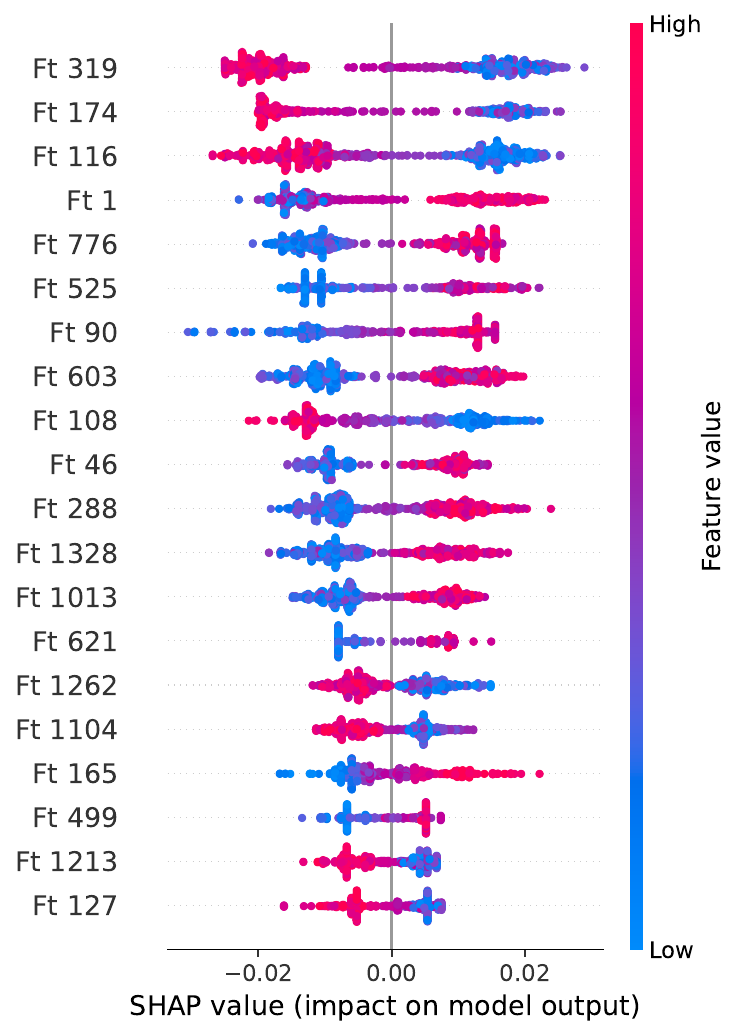}
        \caption{}
        \label{fig:shap_kidneyc}
    \end{subfigure}
    \hfill
    \begin{subfigure}[t]{0.19\textwidth}
        \centering
        \includegraphics[width=\linewidth]{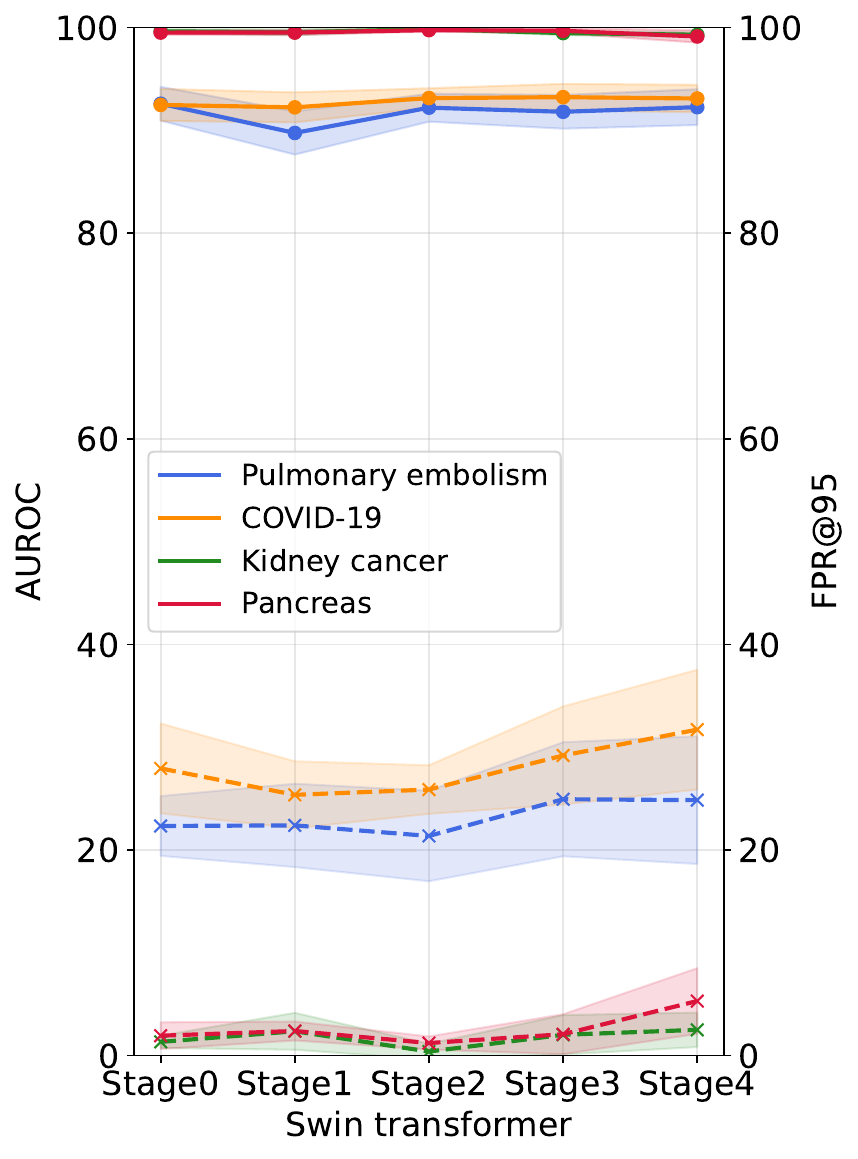}
        \caption{}
        \label{fig:ablation_stage}
    \end{subfigure}
    \caption{Supporting analyses of our approach: (a) t-SNE visualization shows ID/OOD separation in encoder representations (the convex hull denotes extent of ID), (b,c) SHAP analysis highlighting feature importance and interpretability in RSNA-Pulmonary Embolism and KiTS-23 kidney cancer datasets respectively, and (d) Stage-wise ablation shows individual RF-Deep performance from Swin Transformer encoder features.}
    \label{fig:ablations}
\end{figure*}

Our results demonstrate that RF-Deep provides a robust scan-level OOD detection for lung cancer segmentation, outperforming established approaches including MaxSoftmax and MaxLogits (Table~\ref{tab:my-table}). SimMIM-pretrained encoder features, paired with RF, consistently yielded higher AUROC and lower FPR95 on our lung cancer segmentation task, with large margin gains in the abdominal cohorts, achieving near-zero FPR95 on kidney cancer and sub-1\% FPR95 on healthy pancreas. Fig.~\ref{fig:rf_deep_outputs} visually shows RF-Deep's ability to detect scans from completely different anatomical sites and different disease cases, with some limitations. In pulmonary embolism cases, we observed that the segmentation model often highlights tumor-like structures rather than the emboli; nevertheless, our contextual features enable RF-Deep to correctly classify these scans as OOD most of the time, surpassing non-contextual approaches. 

Additionally, we performed t-SNE clustering of the features~\cite{masarczyk2024tunnel} across all cohorts (Fig.~\ref{fig:tsne}) and observed partially distinct clusters for in-distribution (ID) and OOD cohorts, making them favourably separable with the random forest classifier. Fig.~\ref{fig:shap_rsna} and Fig.~\ref{fig:shap_kidneyc} show the SHAP analysis on the RF-Deep providing partial interpretability in identifying features that are most influential in classifying a scan as OOD. Finally, stage-wise ablation (Fig.~\ref{fig:ablation_stage}), wherein only features corresponding to a particular stage of the transformer were used for training RF-Deep, demonstrated that mid- and early-deep level Swin Transformer features are most effective, aligning with our SHAP findings.

\section{Conclusion}

We performed a comprehensive evaluation of different OOD detection methods applied to the clinical task of lung cancer segmentation. Our analysis showed that, even though segmentation models can predict tumors incorrectly at times, RF-Deep, built on encoder features from our fine-tuned model with a random forest, consistently outperforms established methods. The approach is lightweight, interpretable, and improves the reliability of segmentation in both near- and far-OOD clinical scenarios. Future work involves scaling to more disease sites, understanding multi-class behavior, and evaluating generalization across other backbone pretraining strategies.

\newpage
\acknowledgments 
 
This research was partially supported by NCI R01CA258821 and Memorial Sloan Kettering (MSK) Cancer Center Support Grant/Core Grant NCI P30CA008748.

\bibliography{references,references_ood} 

@article{zwanenburg2020image,
  title={The image biomarker standardization initiative: standardized quantitative radiomics for high-throughput image-based phenotyping},
  author={Zwanenburg, Alex and Valli{\`e}res, Martin and Abdalah, Mahmoud A and Aerts, Hugo JWL and Andrearczyk, Vincent and Apte, Aditya and Ashrafinia, Saeed and Bakas, Spyridon and Beukinga, Roelof J and Boellaard, Ronald and others},
  journal={Radiology},
  volume={295},
  number={2},
  pages={328--338},
  year={2020},
  publisher={Radiological Society of North America}
}

@article{deasy2003cerr,
  title={\textsc{CERR}: a computational environment for radiotherapy research},
  author={Deasy, Joseph O and Blanco, Angel I and Clark, Vanessa H},
  journal={Medical physics},
  volume={30},
  number={5},
  pages={979--985},
  year={2003},
  publisher={Wiley Online Library}
}

@article{apte2018extension,
  title={Extension of \textsc{CERR} for computational radiomics: A comprehensive \textsc{MATLAB} platform for reproducible radiomics research},
  author={Apte, Aditya P and Iyer, Aditi and Crispin-Ortuzar, Mireia and Pandya, Rutu and Van Dijk, Lisanne V and Spezi, Emiliano and Thor, Maria and Um, Hyemin and Veeraraghavan, Harini and Oh, Jung Hun and others},
  journal={Medical physics},
  volume={45},
  number={8},
  pages={3713--3720},
  year={2018},
  publisher={Wiley Online Library}
}

@article{yeung2023calibrating,
  title={Calibrating the dice loss to handle neural network overconfidence for biomedical image segmentation},
  author={Yeung, Michael and Rundo, Leonardo and Nan, Yang and Sala, Evis and Sch{\"o}nlieb, Carola-Bibiane and Yang, Guang},
  journal={Journal of Digital Imaging},
  volume={36},
  number={2},
  pages={739--752},
  year={2023},
  publisher={Springer}
}

@misc{heller2023kits21,
      title={The \textsc{KiTS}21 Challenge: Automatic segmentation of kidneys, renal tumors, and renal cysts in corticomedullary-phase CT}, 
      author={Nicholas Heller and Fabian Isensee and Dasha Trofimova and Resha Tejpaul and Zhongchen Zhao and Huai Chen and Lisheng Wang and Alex Golts and Daniel Khapun and Daniel Shats and Yoel Shoshan and Flora Gilboa-Solomon and Yasmeen George and Xi Yang and Jianpeng Zhang and Jing Zhang and Yong Xia and Mengran Wu and Zhiyang Liu and Ed Walczak and Sean McSweeney and Ranveer Vasdev and Chris Hornung and Rafat Solaiman and Jamee Schoephoerster and Bailey Abernathy and David Wu and Safa Abdulkadir and Ben Byun and Justice Spriggs and Griffin Struyk and Alexandra Austin and Ben Simpson and Michael Hagstrom and Sierra Virnig and John French and Nitin Venkatesh and Sarah Chan and Keenan Moore and Anna Jacobsen and Susan Austin and Mark Austin and Subodh Regmi and Nikolaos Papanikolopoulos and Christopher Weight},
      year={2023},
      eprint={2307.01984},
      archivePrefix={arXiv},
      primaryClass={cs.CV}
}

@misc{ricord2021covid,
  doi = {10.7937/31V8-4A40},
  url = {https://www.cancerimagingarchive.net/collection/midrc-ricord-1b/},
  author = {Tsai,  Emily B. and Simpson,  Scott and Lungren,  Matthew P. and Hershman,  Michelle and Roshkovan,  Leonid and Colak,  Errol and Erickson,  Bradley J. and Shih,  George and Stein,  Anouk and Kalpathy-Cramer,  Jayashree and Shen,  Jody and Hafez,  Mona A.F. and John,  Susan and Rajiah,  Prabhakar and Pogatchnik,  Brian P. and Mongan,  John Thomas and Altinmakas,  Emre and Ranschaert,  Erik and Kitamura,  Felipe Campos and Topff,  Laurens and Moy,  Linda and Kanne,  Jeffrey P. and Wu,  Carol},
  title = {Medical Imaging Data Resource Center (\textsc{MIDRC}) - \textsc{RSNA} International \textsc{COVID} Open Research Database (\textsc{RICORD}) Release 1b - Chest CT Covid- },
  publisher = {The Cancer Imaging Archive},
  year = {2021},
  copyright = {Creative Commons Attribution Non Commercial 4.0 International}
}

@article{gu2024build,
  title={How to build the best medical image segmentation algorithm using foundation models: a comprehensive empirical study with segment anything model},
  author={Gu, Hanxue and Dong, Haoyu and Yang, Jichen and Mazurowski, Maciej A},
  journal={arXiv preprint arXiv:2404.09957},
  year={2024}
}

@article{Willemink2022,
	title        = {Toward Foundational Deep Learning Models for Medical Imaging in the New Era of Transformer Networks},
	author       = {Willemink, M.J. and Roth, R.R and Sandfort, V},
	year         = 2022,
	journal      = {Radiol Artif Intell},
	volume       = 4,
	number       = 6
}

@article{masarczyk2024tunnel,
	title        = {The tunnel effect: Building data representations in deep neural networks},
	author       = {Masarczyk, Wojciech and Ostaszewski, Mateusz and Imani, Ehsan and Pascanu, Razvan and Mi{\l}o{\'s}, Piotr and Trzcinski, Tomasz},
	year         = 2024,
	journal      = {Advances in Neural Information Processing Systems},
	volume       = 36
}

@article{aerts2015data,
	title        = {Data from NSCLC-radiomics},
	author       = {Aerts, HJWL and Velazquez, E Rios and Leijenaar, RT and Parmar, Chintan and Grossmann, Patrick and Cavalho, S and Bussink, Johan and Monshouwer, Ren{\'e} and Haibe-Kains, Benjamin and Rietveld, Derek and others},
	year         = 2015,
	journal      = {The cancer imaging archive}
}

@article{bakr2017data,
	title        = {Data for NSCLC radiogenomics collection},
	author       = {Bakr, Shaimaa and Gevaert, Olivier and Echegaray, Sebastian and Ayers, Kelsey and Zhou, Mu and Shafiq, Majid and Zheng, Hong and Zhang, Weiruo and Leung, Ann and Kadoch, Michael and others},
	year         = 2017,
	journal      = {The Cancer Imaging Archive},
	volume       = 10,
	pages        = {K9}
}

@inproceedings{xie2022simmim,
	title        = {Simmim: A simple framework for masked image modeling},
	author       = {Xie, Zhenda and Zhang, Zheng and Cao, Yue and Lin, Yutong and Bao, Jianmin and Yao, Zhuliang and Dai, Qi and Hu, Han},
	year         = 2022,
	booktitle    = {Proceedings of the IEEE/CVF conference on computer vision and pattern recognition},
	pages        = {9653--9663}
}

@article{colak2021rsna,
	title        = {The RSNA pulmonary embolism CT dataset},
	author       = {Colak, Errol and Kitamura, Felipe C and Hobbs, Stephen B and Wu, Carol C and Lungren, Matthew P and Prevedello, Luciano M and Kalpathy-Cramer, Jayashree and Ball, Robyn L and Shih, George and Stein, Anouk and others},
	year         = 2021,
	journal      = {Radiology: Artificial Intelligence},
	publisher    = {Radiological Society of North America},
	volume       = 3,
	number       = 2,
	pages        = {e200254}
}

@inproceedings{liu2021swin,
	title        = {Swin transformer: Hierarchical vision transformer using shifted windows},
	author       = {Liu, Ze and Lin, Yutong and Cao, Yue and Hu, Han and Wei, Yixuan and Zhang, Zheng and Lin, Stephen and Guo, Baining},
	year         = 2021,
	booktitle    = {Proceedings of the IEEE/CVF international conference on computer vision},
	pages        = {10012--10022}
}

@inproceedings{jiang2022self,
	title        = {Self-supervised 3D anatomy segmentation using self-distilled masked image transformer (\textsc{SMIT})},
	author       = {Jiang, Jue and Tyagi, Neelam and Tringale, Kathryn and Crane, Christopher and Veeraraghavan, Harini},
	year         = 2022,
	booktitle    = {International Conference on Medical Image Computing and Computer-Assisted Intervention},
	pages        = {556--566},
	organization = {Springer}
}

@inproceedings{ronneberger2015u,
	title        = {U-net: Convolutional networks for biomedical image segmentation},
	author       = {Ronneberger, Olaf and Fischer, Philipp and Brox, Thomas},
	year         = 2015,
	booktitle    = {International Conference on Medical Image Computing and Computer-Assisted Intervention},
	pages        = {234--241},
	organization = {Springer}
}

@article{paszke2019pytorch,
	title        = {Pytorch: An imperative style, high-performance deep learning library},
	author       = {Paszke, Adam and Gross, Sam and Massa, Francisco and Lerer, Adam and Bradbury, James and Chanan, Gregory and Killeen, Trevor and Lin, Zeming and Gimelshein, Natalia and Antiga, Luca and others},
	year         = 2019,
	journal      = {Advances in neural information processing systems},
	volume       = 32
}

@inproceedings{YanWACV2023,
	title        = {Representation Recovering for Self-Supervised Pre-training on Medical Images},
	author       = {Yan, Xiangyi and Naushad, Junayed and Sun, Shanlin and Han, Kun and Tang, Hao and Kong, Deying and Ma, Haoyu and You, Chenyu and Xie, Xiaohui},
	year         = 2023,
	booktitle    = {2023 IEEE/CVF Winter Conference on Applications of Computer Vision (WACV)},
	volume       = {},
	number       = {},
	pages        = {2684--2694}
}

@inproceedings{NguyenAAAI2023,
	title        = {Joint self-supervised image-volume representation learning with intra-inter contrastive clustering},
	author       = {Nguyen, Duy M. H. and Nguyen, Hoang and Mai, Truong T. N. and Cao, Tri and Nguyen, Binh T. and Ho, Nhat and Swoboda, Paul and Albarqouni, Shadi and Xie, Pengtao and Sonntag, Daniel},
	year         = 2023,
	booktitle    = {Proc. 37th AAAI},
	publisher    = {AAAI Press},
	isbn         = {978-1-57735-880-0},
	articleno    = 1618,
	numpages     = 10
}

@article{cardoso2022monai,
	title        = {Monai: An open-source framework for deep learning in healthcare},
	author       = {Cardoso, M Jorge and Li, Wenqi and Brown, Richard and Ma, Nic and Kerfoot, Eric and Wang, Yiheng and Murrey, Benjamin and Myronenko, Andriy and Zhao, Can and Yang, Dong and others},
	year         = 2022,
	journal      = {arXiv preprint arXiv:2211.02701}
}

@inproceedings{roth2015deeporgan,
	title        = {Deeporgan: Multi-level deep convolutional networks for automated pancreas segmentation},
	author       = {Roth, Holger R and Lu, Le and Farag, Amal and Shin, Hoo-Chang and Liu, Jiamin and Turkbey, Evrim B and Summers, Ronald M},
	year         = 2015,
	booktitle    = {Medical Image Computing and Computer-Assisted Intervention--MICCAI 2015: 18th International Conference, Munich, Germany, October 5-9, 2015, Proceedings, Part I 18},
	pages        = {556--564},
	organization = {Springer}
}

@article{Qayyum2023BHI,
	title        = {Two-Stage Self-Supervised Contrastive Learning Aided Transformer for Real-Time Medical Image Segmentation},
	author       = {Qayyum, Abdul and Razzak, Imran and Mazher, Moona and Khan, Tariq and Ding, Weiping and Niederer, Steven},
	year         = 2023,
	journal      = {IEEE Journal of Biomedical and Health Informatics},
	pages        = {1--10}
}

@article{hendrycks2019anomalyseg,
  title={Scaling Out-of-Distribution Detection for Real-World Settings},
  author={Hendrycks, Dan and Basart, Steven and Mazeika, Mantas and Zou, Andy and Kwon, Joe and Mostajabi, Mohammadreza and Steinhardt, Jacob and Song, Dawn},
  journal={ICML},
  year={2022}
}

@article{konz2024radmetricmedicalimage,
      title={RaD: A Metric for Medical Image Distribution Comparison in Out-of-Domain Detection and Other Applications}, 
      author={Nicholas Konz and Yuwen Chen and Hanxue Gu and Haoyu Dong and Yaqian Chen and Maciej A. Mazurowski},
      year={2024},
      eprint={2412.01496},
      archivePrefix={arXiv},
      primaryClass={cs.CV},
      url={https://arxiv.org/abs/2412.01496}, 
}

@inproceedings{davis2006relationship,
  title={The relationship between Precision-Recall and ROC curves},
  author={Davis, Jesse and Goadrich, Mark},
  booktitle={Proceedings of the 23rd international conference on Machine learning},
  pages={233--240},
  year={2006}
}

@article{liang2017enhancing,
  title={Enhancing the reliability of out-of-distribution image detection in neural networks},
  author={Liang, Shiyu and Li, Yixuan and Srikant, Rayadurgam},
  journal={arXiv preprint arXiv:1706.02690},
  year={2017}
}

@article{mehrtash2020confidence,
  title={Confidence calibration and predictive uncertainty estimation for deep medical image segmentation},
  author={Mehrtash, Alireza and Wells, William M and Tempany, Clare M and Abolmaesumi, Purang and Kapur, Tina},
  journal={IEEE transactions on medical imaging},
  volume={39},
  number={12},
  pages={3868--3878},
  year={2020},
  publisher={IEEE}
}

@article{hendrycks2018deep,
  title={Deep anomaly detection with outlier exposure},
  author={Hendrycks, Dan and Mazeika, Mantas and Dietterich, Thomas},
  journal={arXiv preprint arXiv:1812.04606},
  year={2018}
}

@article{pinaya2022unsupervised,
  title={Unsupervised brain imaging 3D anomaly detection and segmentation with transformers},
  author={Pinaya, Walter HL and Tudosiu, Petru-Daniel and Gray, Robert and Rees, Geraint and Nachev, Parashkev and Ourselin, S{\'e}bastien and Cardoso, M Jorge},
  journal={Medical Image Analysis},
  volume={79},
  pages={102475},
  year={2022},
  publisher={Elsevier}
}

@article{vasiliuk2023limitations,
  title={Limitations of out-of-distribution detection in 3d medical image segmentation},
  author={Vasiliuk, Anton and Frolova, Daria and Belyaev, Mikhail and Shirokikh, Boris},
  journal={Journal of Imaging},
  volume={9},
  number={9},
  pages={191},
  year={2023},
  publisher={MDPI}
}

@article{jiang2024self,
  title = {Self‐supervised learning improves robustness of deep learning lung tumor segmentation models to CT imaging differences},
  ISSN = {2473-4209},
  url = {http://dx.doi.org/10.1002/mp.17541},
  DOI = {10.1002/mp.17541},
  journal = {Medical Physics},
  publisher = {Wiley},
  author = {Jiang,  Jue and Rangnekar,  Aneesh and Veeraraghavan,  Harini},
  year = {2024},
  month = dec 
}

@article{zimmerer2022mood,
  title={Mood 2020: A public benchmark for out-of-distribution detection and localization on medical images},
  author={Zimmerer, David and Full, Peter M and Isensee, Fabian and J{\"a}ger, Paul and Adler, Tim and Petersen, Jens and K{\"o}hler, Gregor and Ross, Tobias and Reinke, Annika and Kascenas, Antanas and others},
  journal={IEEE Transactions on Medical Imaging},
  volume={41},
  number={10},
  pages={2728--2738},
  year={2022},
  publisher={IEEE}
}

@inproceedings{graham2022transformer,
  title={Transformer-based out-of-distribution detection for clinically safe segmentation},
  author={Graham, Mark S and Tudosiu, Petru-Daniel and Wright, Paul and Pinaya, Walter Hugo Lopez and Jean-Marie, U and Mah, Yee H and Teo, James T and Jager, Rolf and Werring, David and Nachev, Parashkev and others},
  booktitle={International Conference on Medical Imaging with Deep Learning},
  pages={457--476},
  year={2022},
  organization={PMLR}
}

@article{hendrycks17baseline,
  author    = {Dan Hendrycks and Kevin Gimpel},
  title     = {A Baseline for Detecting Misclassified and Out-of-Distribution Examples in Neural Networks},
  journal = {Proceedings of International Conference on Learning Representations},
  year = {2017},
}

@inproceedings{hendrycks2022logits,
  title={Scaling Out-of-Distribution Detection for Real-World Settings},
  author={Dan Hendrycks and Steven Basart and Mantas Mazeika and Andy Zou and Mohammadreza Mostajabi and Jacob Steinhardt and Dawn Xiaodong Song},
  booktitle={International Conference on Machine Learning},
  year={2022},
  url={https://api.semanticscholar.org/CorpusID:227407829}
}

@article{liu2020energy,
  title={Energy-based out-of-distribution detection},
  author={Liu, Weitang and Wang, Xiaoyun and Owens, John and Li, Yixuan},
  journal={Advances in neural information processing systems},
  volume={33},
  pages={21464--21475},
  year={2020}
}
\bibliographystyle{spiebib} 

\end{document}